\documentclass[aps,prl,nopacs,floatfix,superscriptaddress,twocolumn]{revtex4-1}

\usepackage{graphicx}
\usepackage{dcolumn}
\usepackage{bm}

\usepackage[colorlinks=true,linkcolor=blue,citecolor=red, linktocpage=true,breaklinks=true]{hyperref}

\usepackage{dsfont}

\usepackage[space]{grffile}
\usepackage{latexsym}
\usepackage{textcomp}
\usepackage{multirow,booktabs}
\usepackage{amsfonts,amsmath,amssymb}
\usepackage{url}
\usepackage[utf8]{inputenc}
\usepackage[english]{babel}
\usepackage{color,verbatim}
\usepackage{psfrag}
\usepackage{tcolorbox}
\usepackage{times}
\usepackage{enumerate,mdwlist}

\newcommand{\ba}{\begin{eqnarray}}
\newcommand{\be}{\begin{equation}}
\newcommand{\ee}{\end{equation}}
\newcommand{\beq}{\begin{equation}}
\newcommand{\eeq}{  \end{equation}}
\newcommand{\bea}{\begin{eqnarray}}
\newcommand{\eea}{  \end{eqnarray}}
\newcommand{\g}{\textrm{gen}}

\newcommand{\ea}{\end{eqnarray}}
\newcommand{\ban}{\begin{eqnarray*}}
\newcommand{\ean}{\end{eqnarray*}}

\newcommand{\tr}{\operatorname{tr}}

\newcommand{\ket}[1]{\vert #1 \rangle}
\newcommand{\bra}[1]{\langle #1 \vert}

\newcommand{\ie}{{\it{i.e.}~}}

\newcommand{\JM}{\mathrm{C}}

\definecolor{ivan}{rgb}{0.7,0,0.7}
 
\usepackage{subcaption}

\newcommand{\ab}{\mathbf{a}}

\begin{document}

\title{All sets of incompatible measurements give an advantage in quantum state discrimination}

\author{Paul Skrzypczyk}
\affiliation{H. H. Wills Physics Laboratory, University of Bristol, Tyndall Avenue, Bristol, BS8 1TL, United Kingdom}%
\email{paul.skrzypczyk@bristol.ac.uk}

\author{Ivan \v{S}upi\'{c}}
\affiliation{D{\'{e}}partement de Physique Appliqu\'{e}e, Universit\'{e} de Gen\`{e}ve, 1211 Gen\`{e}ve, Switzerland}
\email{ivan.supic@unige.ch}

\author{Daniel Cavalcanti}
\affiliation{ICFO-Institut de Ciencies Fotoniques,  The Barcelona Institute of Science and Technology,  08860 Castelldefels (Barcelona),  Spain}
 \email{daniel.cavalcanti@icfo.eu}

\date{\today}

\begin{abstract}
Some quantum measurements can not be performed simultaneously, \ie they are incompatible. Here we show that every set of incompatible measurements provides an advantage over compatible ones in a suitably chosen quantum state discrimination task. This is proven by showing that the Robustness of Incompatibility, a quantifier of how much noise a set of measurements tolerates before becoming compatible, has an operational interpretation as the advantage in an optimally chosen discrimination task. We also show that if we take a resource-theory perspective of measurement incompatibility, then the guessing probability in discrimination tasks of this type forms a complete set of monotones that completely characterize the partial order in the resource theory. Finally, we make use of previously known relations between measurement incompatibility and Einstein-Podolsky-Rosen steering to also relate the later with quantum state discrimination.

\end{abstract}

\maketitle
\section{Introduction}
In quantum mechanics, observables described by non-commuting operators satisfy an uncertainty relation, which implies that we can not acquire precise information about them simultaneously \cite{Robertson29}. First thought to be a limitation, recent advances in quantum information theory have demonstrated that this feature is behind several applications, such as the security of quantum key distribution \cite{cripto_review}, and nonlocality based (or device-independent) applications \cite{NL_review}.

Commutation is well defined for sharp (von Neumanm) measurements. However, a more refined notion of measurement incompatibility is needed for general measurements described by positive-operator-value-measures (POVMs) \cite{krausbook}. This is captured by the idea of joint measurability \cite{IncompatRev}. Suppose a set of measurements $\{\mathbb{M}_x\}_x$ labeled by $x=1,\ldots,m$, each described by measurement operators $M_{a|x}$ ($M_{a|x}\geq 0$, $\sum_a M_{a|x}=\openone ~ \forall ~a,x$), where $a=1,\ldots,o$ labels each of the measurement outcomes. This set is said to be jointly measurable (or compatible) if there exists a `parent' measurement $\mathbb{G}$ with measurement operators $G_\lambda$, and conditional probability distributions $p(a|x,\lambda)$, such that
\be\label{eq: JM}
M_{a|x}=\sum_\lambda p(a|x,\lambda) G_\lambda \quad \forall ~a,x.
\ee
Otherwise the set is said to be incompatible. This definition can be interpreted as follows: if \eqref{eq: JM} holds, all measurements $\mathbb{M}_x$ can be performed jointly, by the implementation of the single measurement $\mathbb{G}$ and a probabilistic classical post-processing defined by the weights $p(a|x,\lambda)$.

 Here we give an operational interpretation of measurement incompatibility in terms of quantum state discrimination: we show that a set of measurements is incompatible if and only if they provide an advantage over compatible ones in a quantum state discrimination (QSD) task with multiple ensembles of states. Moreover, we also show that the advantage of an optimally chosen QSD task is quantified exactly by the robustness of incompatibility of the set, a previously proposed quantifier of measurement incompatibility \cite{Uola15}. This result fits within a number of results recently obtained which have linked robustness-based quantifiers with advantages in suitably chosen discrimination games \cite{piani2015, napoli2016,takagi2018,bae2018,SkrLin18} 

\section{Incompatibility and Advantage in Quantum State Discrimination}
We consider the following two-party QSD task \cite{StDisc}: Bob can prepare different ensembles $\{\mathcal{E}_y\}_y$ ($y = 1,\ldots,n$) of quantum states $\mathcal{E}_y = \{\rho_{b|y},q(b|y)\}_b$,  for $b = 1,\ldots,p$. At each round of the protocol, Bob chooses one of the ensembles $y$ with probability $q(y)$ and sends Alice his choice $y$, and the state prepared $\rho_{b|y}$, which occurs with probability $q(b|y)$. Upon receiving $y$ and $\rho_{b|y}$, Alice's goal is to identify which state she was sent, \ie to correctly identify $b$.

We will consider playing this game in two different scenarios. In the first scenario, Alice has access to a fixed set of incompatible measurements $\{\mathbb{M}_x\}_x$ in order to play. We consider the most general probabilistic strategies assuming that the only way Alice can interact with the system is through her fixed measuring device. In particular, we allow any strategy consisting of the following \footnote{Note that a more general class of strategies would allow for a pre-processing of the state also, i.e. the application of an arbitrary quantum instrument (collection of completely positive maps that sum to a trace-preserving channel). Here we do not give Alice such capabilities, but demand that the only Alice interact directly with the quantum system sent to her is through the measuring device corresponding to the incompatible measurements}: After receiving the state and the value of $y$, Alice makes use of a random variable $\mu$ to perform the measurement $\mathbb{M}_x$, with probability $p(x|y,\mu)$. After receiving outcome $a$ she makes a guess of the value of $b$, according to $p(g|a,y,\mu)$. Optimizing over all strategies, we can quantify how well Alice does in this game by evaluating the average probability of correctly identifying $b$, i.e.
\begin{multline}\label{e:Pg}
    P_g(\{\mathcal{E}_y\},\{\mathbb{M}_x\}) = \max_\mathcal{S} \sum_{byaxg\mu}q(b,y)p(\mu)p(x|y,\mu)\\ \times \tr[\rho_{b|y}M_{a|x}]p(g|a,y,\mu) \delta_{g,b}
\end{multline}
where the maximization is over strategies $\mathcal{S} = \{p(\mu), p(x|y,\mu), p(g|a,y,\mu)\}$, and we have written $q(b,y) = q(y)q(b|y)$.

We will contrast this to a scenario where in any given run of the game Alice can only perform a single measurement (although we will allow once again the possibility of using randomness to mix over different fixed measurements in different runs of the game). In particular, we consider measurements $\mathbb{G}_\nu = \{G_{a|\nu}\}_a$, and allow for the most general strategy using any such measurements. Crucially now, since Alice can only perform a single measurement, the side-information of $y$ can only be used to implement a classical post-processing of this measurement. The net effect is equivalent to Alice only being able to perform a set of compatible measurements, those achieved by the `parent' measurements $\mathbb{G}_\nu$. In this case the success probability is given by
\begin{multline}\label{e:psuc C}
    P_g^{\JM}(\{\mathcal{E}_y\}) = \max_\mathcal{T} \sum_{bya\nu g} q(b,y) p(\nu) \\ \times \tr[\rho_{b|y}G_{a|\nu}]p(g|a,y,\nu)\delta_{g,b}
\end{multline}
where the maximization is over all strategies $\mathcal{T} = \{p(\nu), \mathbb{G}_\nu, p(g|a,y,\nu)\}$.

We are primarily interested in the \emph{advantage} that is offered by a set of incompatible measurements $\{\mathbb{M}_x\}_x$ in any such QSD game. In particular, we are interested in the biggest relative increase in guessing probability that can be obtained by the set of measurements $\{\mathbb{M}_x\}_x$ compared to having access to only single measurements, among all possible ensembles, i.e. 
\begin{equation} \label{e: advantage}
    \max_{\{\mathcal{E}_y\}} \frac{P_g(\{\mathcal{E}_y\},\{\mathbb{M}_x\})}{P_g^{\JM}(\{\mathcal{E}_y\})}
\end{equation}

The main result of this Letter is to show that this quantity is completely characterised by the Robustness of Incompatibility (RoI) of the measurements $ I_R(\{\mathbb{M}_x\})$ as
\begin{equation}\label{e: main}
   1+ I_R(\{\mathbb{M}_x\})= \max_{\{\mathcal{E}_y\}} \frac{P_g(\{\mathcal{E}_y\},\{\mathbb{M}_x\})}{P_g^{\JM}(\{\mathcal{E}_y\})}.
\end{equation}
The Robustness of Incompability $ I_R(\{\mathbb{M}_x\})$ is defined as the minimal amount of `noise' that needs to be added to the set of measurements $\{\mathbb{M}_x\}_x$ before they become compatible \cite{Uola15}. Here, by `noise', we mean that we mix the set of measurements with another, arbitrary, set of measurements $\{\mathbb{N}_x\}_x$, (of the same size, and with the same number of outcomes), in order to make the mixture compatible. Formally, 
\begin{align}\label{e: IR}
    I_R(\{\mathbb{M}_x\}) = \min &\quad r \\
    \text{s.t.} &\quad \frac{M_{a|x} + rN_{a|x}}{1+r} = \sum_\lambda p(a|x,\lambda)G_\lambda \nonumber\\
    &\quad N_{a|x} \geq 0, \quad \sum_a N_{a|x} = \openone, \nonumber \\ 
    &\quad p(a|x,\lambda) \geq 0,\quad \sum_a p(a|x,\lambda) = 1, \nonumber \\
    &\quad G_\lambda \geq 0, \quad \sum_\lambda G_\lambda = \openone \nonumber
\end{align}
where the minimisation is over $r$, $\{\mathbb{N}_x\}_x$ (where $\mathbb{N}_x = \{N_{a|x}\}_a$), $\mathbb{G} = \{G_\lambda\}_\lambda$ and $\{p(a|x,\lambda)\}_{a,x,\lambda}$, and all constraints are understood to hold for all values of $a$, $x$, or $\lambda$, as appropriate. 

The RoI has a number of desirable properties: 
\begin{enumerate}[(i)]
    \item It is faithful: $I_R(\{\mathbb{M}_x\}) = 0$ if and only if the set of measurements $\{M_x\}_x$ is incompatible;

\item It is convex: If the set of measurements $\{\mathbb{M}_x\}_x$ is a convex combination of two other sets of measurements, i.e. for all $x$, $\mathbb{M}_x = p\mathbb{M}^{(1)}_x + (1-p)\mathbb{M}^{(2)}_x$, for some $p > 0$, and for valid sets of measurements $\{\mathbb{M}^{(1)}_x\}_x$ and $\{\mathbb{M}^{(2)}_x\}_x$, then
\begin{equation}
    I_R(\{\mathbb{M}_x\}) \leq pI_R(\{\mathbb{M}^{(1)}_x\}) + (1-p)I_R(\{\mathbb{M}^{(2)}_x\})
\end{equation}
\item  It is non-increasing under post-processing of the measurements. That is, if we simulate a new set of measurements $\{\mathbb{M}'_y\}_y$ using $\{\mathbb{M}_x\}_x$, such that 
\begin{equation}\label{e:sim}
    M'_{b|y} = \sum_{a,x,\mu}p(\mu) p(x|y,\mu) p(b|a,y,\mu)M_{a|x}
\end{equation}
where $p(\mu)$, $p(x|y,\mu)$ and $p(b|a,y,\mu)$ are arbitrary sets of probability distributions, then 
\begin{equation}
    I_R(\{\mathbb{M}'_y\}) \leq I_R(\{\mathbb{M}_x\}).
\end{equation}
\end{enumerate}
Due to \eqref{e: main}, the  properties (i) -- (iii) are also satisfied by the  advantage $\eqref{e: advantage}$. In particular, due to (i), a set of measurements $\{\mathbb{M}_x\}_x$ provides an advantage over compatible measurements if and only if  $I_R(\{\mathbb{M}_x\})>0$. 

Another interesting consequence of \eqref{e: main} is that it gives an efficient way of computing the advantage \eqref{e: advantage}. This is because the RoI can be shown to be expressed explicitly as the following semi-definite program (SDP):
\begin{align} \nonumber
    1 + I_R(\{\mathbb{M}_x\}) = \min_{s,\{\tilde{G}_\mathbf{a}\}}&\quad s\\ \label{primal}
    \textrm{s.t.} &\quad\sum_\mathbf{a} D_\mathbf{a}(a|x)\tilde{G}_\mathbf{a} \geq M_{a|x}\\
    &\quad \sum_\mathbf{a}\tilde{G}_\mathbf{a} = s\openone, \quad \tilde{G}_\mathbf{a} \geq  0 \nonumber 
\end{align}
where $\mathbf{a} = \mathrm{a}_1 \mathrm{a}_2 \cdots \mathrm{a}_n$ is a string, which can be throught of as a list of `results', one for each measurement, $D_\mathbf{a}(a|x) = \delta_{a,\mathrm{a}_x}$ are deterministic probability distributions, whereby $a = \mathrm{a}_x$ with certainty, and $\tilde{\mathbb{G}} = \{\tilde{G}_\mathbf{a}\}_\mathbf{a}$ is a super-normalised parent POVM. The derivation of this SDP formulation can be found in the appendix. 

Let us now sketch the proof of our main result (we leave the full proof for the appendix). Consider that the solution of \eqref{e: IR} is attained by $N_{a|x}^*$, $G^*_{\lambda}$, and $p^*(a|x,\lambda)$, which means that 
\begin{align}
    \frac{M_{a|x} + I_R(\{\mathbb{M}_x\})N_{a|x}^*}{1+I_R(\{\mathbb{M}_x\})} = \sum_\lambda p^*(a|x,\lambda)G_\lambda^* .
\end{align}
Since $I_R(\{\mathbb{M}_x\})\geq0$ and $N_{a|x}^*\geq0$, we have that
\begin{align*}
[1+I_R(\{\mathbb{M}_x\})]\sum_\lambda p^*(a|x,\lambda)G_\lambda^* \geq M_{a|x} \qquad \forall a,x.
\end{align*}
Multiplying both sides of this expression by $\rho_{b|y}$, the probabilities appearing the QSD game, taking the trace and applying the correct maximisations, we end up proving that 
\begin{equation}\label{e: upper bound}
1+I_R(\{\mathbb{M}_x\}) \geq \frac{P_g(\{\mathcal{E}_y\},\{\mathbb{M}_x\})}{P_g^{\JM}(\{\mathcal{E}_y\})}.
\end{equation}
This expression is interesting by itself: it states that the RoI of a set of measurements provides an upper bound on the advantage that set provides in \textit{any} QSD game (of the type considered here), defined by the ensembles $\{\mathcal{E}_y\}_y$.

The second part of the proof consists in explicitly showing that for any set $\{\mathbb{M}_x\}_x$ there exists a choice $\{\mathcal{E}_y^*\}_y$ saturating the bound \eqref{e: upper bound}. Such a collection of ensembles can be constructed by using the duality theory of semidefinite programming \cite{boyd2004}. In particular, in the appendix we show that an equivalent formulation of the RoI (the dual formulation) is 
\begin{align} \nonumber
    1 + I_R(\{\mathbb{M}_x\}) = \max_{\{\omega_{ax}\},X}&\quad \tr\sum_{a,x}\omega_{ax}M_{a|x}\\ 
    \textrm{s.t.} &\quad X \geq \sum_{a,x}\omega_{ax}D_\ab(a|x),\\
    &\quad \omega_{ax} \geq 0, \quad \tr X = 1 \nonumber 
\end{align}
Assuming that the maximum is attained by $\{\omega^*_{ax}\}_{ax}$, we can interpret these as unnormalised quantum states, which can be appropriately normalised, and from which we can then define a game through  $\{\mathcal{E}_x^*\}_x$. We show in the appendix that the advantage that $\{\mathbb{M}_x\}_x$ provide in playing this game is precisely $1 + I_R(\{\mathbb{M}_x\})$, which completes the proof. 

To summarise, the above shows that the RoI, which was introduced as a purely geometrical quantifier of incompatibility, in fact has an operational interpretation as the advantage that a set of measurements provides in an optimally chosen QSD game. Moreover, since the RoI is faithful (property (i) above), every set of incompatible measurements gives an advantage in at least one QSD task, and thus this task captures the utility of incompatible measurements.   

\section{Resource theory of Incompatibility}
We now turn to the next result of this Letter, and consider a resource-theory of measurement incompatibility. We will see that this allows us to connect the notion of simulability of one set of measurements by another one, as given in \eqref{e:sim}, with the success probability of these sets in any QSD games considered here. 

In any resource theoretic setting, there are 3 main ingredients \cite{chitambar2018}: (i) a set of free / resourceless objects (ii) a set of expensive / resourceful objects (iii) a set of allowed transformations between objects, which should not be able to create resourceful objects from free objects.  In the present setting, a resource theory of incompatible measurements can easily be formalised: (i) the free objects are the set of all compatible measurements (ii) the resourceful objects are the set of all incompatible measurements (iii) the set of allowed transformations consist of all simulations, i.e. we think of the simulation protocol of \eqref{e:sim} as `transforming' the set of measurements $\{\mathbb{M}_x\}_x$ into the set $\{\mathbb{M}'_y\}_y$. From properties (i) and (iii) of the RoI, we see that any set of compatible measurements cannot be transformed into a set of incompatible ones by measurement simulation, and hence this is a consistent set of allowed transformations.  

Within any resource theory, there is a natural partial order that arises between the objects of the theory: if one object can be transformed into another, then it is `before' it in the partial order. A basic question in any resource theory is then to understand the partial order -- i.e. to find necessary and sufficient conditions which characterise whether one object can be transformed into another or not. Intuitively, objects can only be transformed into other objects which are not more resourceful than themselves, i.e. generalising the idea that the allowed transformations not only cannot create resources from nothing, but cannot increase resources.  

Any function of an object that cannot increase under an allowed transformation is known as a resource \emph{monotone}, and act as witnesses that one object cannot be transformed into another object. In the present setting, property (iii) of the RoI shows that it is a monotone for the resource theory of incompatibility. It is however only a single monotone, and $I_R(\{\mathbb{M}_x\}) > I_R(\{\mathbb{M}_y\})$ does not in general imply that $\{\mathbb{M}_x\}_x$ can simulate $\{\mathbb{M}'_y\}_y$.

In the appendix, inspired by the connection between the RoI and QSD, we prove that \eqref{e:sim} holds, which we will denote simply by $\{\mathbb{M}_{x}\} \succ \{\mathbb{M}'_{y}\}$, if and only if $\{\mathbb{M}_{x}\}$  outperforms $\{\mathbb{M}'_{y}\} $ in every single QSD game of the type considered above, \ie
\begin{multline}\label{e: monotones}
P_g(\{\mathcal{E}_y\},\{\mathbb{M}_{x}\}) \geq P_g(\{\mathcal{E}_y\},\{\mathbb{M}'_y\}) \quad \forall \{\mathcal{E}_y\}_y \\
 \iff \{\mathbb{M}_{x}\} \succ \{\mathbb{M}'_{y}\}
\end{multline} 

Notice that the backward implication ($\Leftarrow$) is natural: if  $\{\mathbb{M}_{x}\}$ can simulate $\{\mathbb{M}'_{y}\}$, then it is obviously contradictory that there is a game where $\{\mathbb{M}'_y\}$ can outperform $\{\mathbb{M}_x\}$. Interestingly, the forward implication ($\Rightarrow$) holds, which proves that the QSD games studied here constitute a complete set of operational monotones that determine if a set of measurements can simulate another. This, in particular, indicates that they capture the resource of incompatibility.     

\section{EPR steering and entanglement-based QSD}
Let us finally describe a connection between the present results and the notion of Einstein-Podolsky-Rosen (EPR) steering \cite{WJD07}. In the EPR steering scenario Alice and Bob share a bipartite quantum state $\rho_{AB}$, onto which Alice applies measurements $\mathbb{M}_x$, leaving Bob's state in the (unnormalised) post-measurement states $\sigma_{a|x}=\tr_A[(M_{a|x}\otimes \openone) \rho_{AB}]$. The set of states  $\{\sigma_{a|x}\}_{a,x}$ -- referred to as an assemblage \cite{Pusey13} -- is said to demonstrate EPR steering if they do not admit  a local-hidden-state (LHS) decomposition of the type $\sigma_{a|x}=\sum_\lambda p(a|x,\lambda) \sigma_\lambda$, where $p(a|x,\lambda)$ are conditional probability distributions and $\sigma_\lambda$ (unnormalised) quantum states \cite{WJD07}. Similarly to the case of incompatibility, the robustness of steerability $S_R(\{\sigma_{a|x}\})$ of $\{\sigma_{a|x}\}_{a,x}$ can be defined as the minimum amount of noise that has to be mixed with each state $\sigma_{a|x}$ from the assemblage, such that it admits a LHS decomposition \cite{PW15}. It is straightforward to see that if $\{\mathbb{M}_x\}_x$ are a compatible set of measurements, then no matter which state $\rho_{AB}$ is used in a steering experiment, all resulting assemblages $\{\sigma_{a|x}\}_{a,x}$ have a LHS decomposition. In the other direction, it also turns out that every set of incompatible measurements has the potential of generating steering \cite{JM_St_Siegen,JM_St_Geneva}. That is, for every set of incompatible measurements there exists bipartite states which demonstrate steering if Alice uses them. 

In what follows we make use of the connection between measurement incompatibility and EPR steering to also connect the latter with QSD and to show that the advantage in the QSD game here can be estimated in the so-called one-sided device-independent paradigm (1SDI) \cite{1SDI} where the set of measurements $\{\mathbb{M}_x\}$ are treated as a black box, such that we don't know the specific measurements made, or the dimension of system they act upon.

In order to accommodate the steering scenario let us describe an entanglement-based variation of the QSD scenario discussed before. Suppose that Bob tells Alice that he is going to measure his part of $\rho_{AB}$ with the measurement $\mathbb{M}_y= \{M_{b|y}\}_b$ (such a measurement can be thought as of performing remote state preparation \cite{Bennett01} of the states $\rho_{b|y}$ of Alice). Once again, Alice's goal is to make a measurement on her system in order to best guess Bob's outcome $b$ (which is equivalent to guessing which state she will receive). 

It was shown in \cite{JM_St_NL} that a 1SDI lower bound can be placed the RoI, 
\begin{equation}
    S_R^c(\{\sigma_{a|x}\})\leq I_R(\{\mathbb{M}
_{x}\}),
\end{equation} 
where $\{\sigma_{a|x}\}_{a,x}$ is an assemblage created by performing the measurements $\{\mathbb{M}_x\}_x$ on any state $\rho_{AB}$, and $S_R^c(\{\sigma_{a|x}\})$ is the \emph{consistent} steering robustness, given by
\begin{align}\label{e: SRc}
    S_R^c(\{\sigma_{a|x}\}) = \min &\quad s \\
    \text{s.t.} &\quad \frac{\sigma_{a|x} + s\omega_{a|x}}{1+s} = \sum_\lambda p(a|x,\lambda)\sigma_\lambda \nonumber\\
    &\quad \omega_{a|x} \geq 0, \quad \sigma_\lambda \geq 0, \nonumber \\ 
    &\quad p(a|x,\lambda) \geq 0,\quad \sum_a p(a|x,\lambda) = 1, \nonumber \\
    &\quad \sum_a \omega_{a|x} = \sum_a \sigma_{a|x} = \sum_\lambda \sigma_\lambda\nonumber
\end{align}
which can be seen as a modification of the steering robustness, with the additional constraint that the `noise' must have the same reduced state as the input assemblage \cite{JM_St_NL}. Moreover, when $\rho_{AB}$ is a pure entangle state (of full Schmidt-rank), then $S_R^c(\{\sigma_{a|x}\}) =  I_R(\{\mathbb{M}
_{x}\})$, i.e. the bound is in fact tight. 

 This means that $1+S_R^c(\{\sigma_{a|x}\})$ provides a 1SDI lower bound on the best advantage that Alice has in guessing $b$ if she measures a set of incompatible measurements instead of a compatible one, and that if Alice and Bob share a pure entangled state, that this bound is in fact tight. 

\section{Conclusions}
In this Letter we have shown that measurement incompatibility, one of the most fundamental features of quantum mechanics, is intrinsically connected the task of discriminating quantum states from collections of ensembles. Our results thus provide an operational interpretation of measurement incompatibility. Moreover it shows that the robustness of incompatibility of a set of measurements is directly related to their usefulness for a natural quantum information game. Finally, we considered a resource theory of measurement incompatibility, and showed that the very same game is intimately related to the simulability of one set of measurements by another, providing (an infinite number of) criteria -- often referred to as monotons -- that collectively constitute necessary and sufficient conditions that must be met for one set of measurements to simulate another. This is similar to a number of other resource theories, where guessing probabilities in all discrimination games of a given type have also been shown to constitute complete criteria for transformations amount objects in the theory \cite{buscemi2016,gour2017,SkrLin18}.

There are a number of natural questions and extensions that we leave for future work. For example, it is interesting to consider partial notions of imcompatibility (i.e. sets of measurements which are pairwise compatible, but not compatible as a complete set), and to ask whether there exist QSD games which characterise the usefulness of such sets. One can also consider generalisations of incompatibility in the other direction, where multiple parent measurements are allowed, and ask similar questions. 

\section{Acknowledgements} PS acknowledges support from the Royal Society through a URF (UHQT). I{\v{S}} acknowledges support from the Swiss National Science Foundation (Starting grant DIAQ). DC acknowledges support from a Ramon y Cajal fellowship, Spanish MINECO (Severo Ochoa SEV-2015-0522), Fundació Privada Cellex and Generalitat de Catalunya (CERCA Program).

\subsection{Note added} While preparing this manuscript we became aware of the following related papers: C. Carmeli, T. Heinosaari, A. Toigo,  arXiv:1812.02985 \cite{Heinosaari18}; R. Uola, et al, arXiv:1812.09216 \cite{Uola18}.

\bibliography{JMvsQSD.bib}

\newpage
\begin{appendix}
\section{APPENDIX}
\section{Incompatibility Robustness -- primal SDP formulation}

In this section we show the equivalence between \eqref{e: IR} and the primal form of the SDP optimization problem \eqref{primal}.
The first constraint can be used to solve for the elements of the `noise' POVM, namely
\begin{equation}
    N_{a|x} = \frac{(1+r)\sum_{\lambda}p(a|x,\lambda)G_{\lambda} - M_{a|x}}{r} \qquad \forall a,x.
\end{equation}
By denoting $s = 1+r$, the positivity of the POVM elements $N_{a|x}$ is then equivalent to
\begin{equation}\label{constraint1}
    s\sum_{\lambda}p(a|x,\lambda)G_{\lambda} \geq M_{a|x}\qquad \forall a,x
\end{equation}
Now note that without loss of generality one can decompose the probabilities $p(a|x,\lambda)$ as a sum of deterministic probabilities, $p(a|x,\lambda) = \sum_\mathbf{a}D_\mathbf{a}(a|x)p(\mathbf{a}|\lambda)$, where $\mathbf{a} = \textrm{a}_1\textrm{a}_2\cdots\textrm{a}_n$ is a string of outcomes (one for each value of x) and $D_{\mathbf{a}}(a|x) = \delta_{a,\mathbf{a}_x}$, i.e. such that $a = \mathrm{a}_x$ with certainty. We can then write
\begin{equation}
    \sum_{\lambda}p(a|x,\lambda)G_{\lambda} = \sum_\mathbf{a} D_\mathbf{a}(a|x)G_\mathbf{a}
\end{equation}
where $G_\mathbf{a} = \sum_\lambda p(a|x,\lambda)G_\lambda$. Each $G_\mathbf{a}$ is positive semidefinite, and they sum to the identity operator, hence they form a valid POVM. This form of parent can be thought of as a canonical parent POVM. 
Finally, we note that we can define $\tilde{G}_\mathbf{a} = s G_\mathbf{a}$, which is a super-normalised POVM, i.e. such that
\begin{equation}\label{constraint2a}
   \tilde{G}_{\mathbf{a}} \geq 0 \quad \forall \mathbf{a}
\end{equation}
and 
\begin{equation}\label{constraint2b}
\sum_{\mathbf{a}}\tilde{G}_{\mathbf{a}} = s\sum_{\mathbf{a}}G_{\mathbf{a}} = s\mathds{1}.
\end{equation}
Gathering the constraints \eqref{constraint1}, \eqref{constraint2a} and \eqref{constraint2b}, one obtains the primal SDP form
\begin{align*} 
    1 + I_R(\{\mathbb{M}_x\}) = \min_{s,\{\tilde{G}_\mathbf{a}\}}&\quad s\\ 
    \textrm{s.t.} &\quad\sum_\mathbf{a} D_\mathbf{a}(a|x)\tilde{G}_\mathbf{a} \geq M_{a|x}\\
    &\quad \sum_\mathbf{a}\tilde{G}_\mathbf{a} = s\openone, \quad \tilde{G}_\mathbf{a} \geq  0
\end{align*}
We see that this is now explicitly in the form of an SDP, since all constraints are linear equalities or inequalities (given that $D_\mathbf{a}(a|x)$ are not variables, but are fixed functions). 

\section{ Incompatibility Robustness -- dual formulation}

In this section we derive the dual SDP formulation of the  RoI. The Lagrangian associated to the primal form of the SDP \eqref{primal} is given by 
\begin{align}\nonumber
    \mathcal{L} &= s + \sum_{a,x}\textrm{tr}\left[\omega_{ax}\left(M_{a|x} - \sum_{\ab}D_{\ab}(a|x)\tilde{G}_\ab\right)\right] - \\  &\qquad-\tr\left[X\left(s\mathds{1} - \sum_\ab\tilde{G}_\ab\right)\right] - \tr\sum_\ab y_\ab\tilde{G}_\ab,
\end{align}
where we have introduced dual variables $\omega_{ax}$ and $y_\ab$, which are taken to be positive-semidefinite for all $a$, $x$ and $\ab$ respectively, and $X$ is an unrestricted dual variable. The constraints on the dual variables are imposed to ensure that the Lagrangian lower bounds the primal objective function whenever the primal constraints are satisfied. By grouping terms, the Lagrangian can be re-expressed as
\begin{align}\nonumber
    \mathcal{L} 
    &= s(1- \tr X) + \tr\sum_{a,x}\omega_{ax}M_{a|x}  + \\ 
    & \qquad + \tr\sum_\ab\tilde{G}_\ab \left[X - \sum_{a,x}\omega_{ax}D_\ab(a|x) - y_\ab\right]
\end{align}
The Lagrangian becomes independent of the primal variables if we restrict to dual variables that satisfy $\tr X = 1$ and $X = \sum_{a,x}\omega_{ax}D_\ab(a|x) + y_\ab$ for all $\ab$. In this case the Lagrangian becomes equal to $\tr\sum_{a,x}\omega_{ax}M_{a|x}$. Hence, the dual form of the SDP reads
\begin{align} \nonumber
    1 + I_R(\{\mathbb{M}_x\}) = \max_{\{\omega_{ax}\},X}&\quad \tr\sum_{a,x}\omega_{ax}M_{a|x}\\ \label{dual}
    \textrm{s.t.} &\quad X \geq \sum_{a,x}\omega_{ax}D_\ab(a|x),\\
    &\quad \omega_{ax} \geq 0, \quad \tr X = 1 \nonumber 
\end{align}
The optimal values of the primal and the dual formulation coincide if  strong duality holds. This is true if there exist a strictly feasible solution of the dual problem (and both problems are finite). An explicit strictly feasible solution is $X = \mathds{1}/d$, $\omega_{ax} = \alpha\mathds{1}$ for any $d$ and $\alpha$ such that $1/nd > \alpha > 0$. The existence of a strictly feasible solution thus ensures the equivalence between the primal and dual SDP formulations. 

\section{Upper bound on the advantage in QSD from the primal SDP}
In this section we show that the RoI for a set of measurements upper bounds the advantage that the set of measurements has in the QSD game defined in the main text, compared to the optimal success which can be achieved with a single measurement. To see this, we start from the original formulation \eqref{e: IR} of the RoI. Let us denote by $G^*_{\lambda}$  and $p^*(a|x,\lambda)$ the optimal parent POVM attaining the minimum. Since the POVM elements of the noise $N_{a|x}$ are positive semi-definite, it follows that
\begin{align}
[1+I_R(\{\mathbb{M}_x\})]\sum_\lambda p^*(a|x,\lambda)G^*_\lambda \geq M_{a|x} \quad \forall a,x.
\end{align}

By taking the trace on both sides with $\rho_{b|y}$, and by multiplying by the appropriate probabilities and summing, this implies that
\begin{multline}\label{sm1}
[1+I_R(\{\mathbb{M}_x\})]\sum_{\substack{\lambda g \mu \\ a bxy}}q(b,y)p(\mu)\tr\left[\rho_{b|y}G^*_{\lambda}\right]\\ \quad\quad\quad\times p^*(a|x,\lambda)p(x|y,\mu)p(g|a,y,\mu)\delta_{b,g} \\
\geq \sum_{\mu abxyg}q(b,y)p(\mu)\tr\left[\rho_{b|y}M_{a|x}\right] \\ \times p(x|y,\mu)p(g|a,y,\mu)\delta_{b,g},
\end{multline}
where $\mathcal{E}_y = \{ q(b|y),\rho_{b|y}\}_b$ represents the ensembles for a QSD game, which occur with probability $q(y)$ (such that $q(y)q(b|y) = q(b,y)$) and $p(g|a,y,\lambda)$ the guessing strategy, as given in \eqref{e:Pg} from the main text. Let us define 
\begin{equation}
p(g|\lambda,y,\mu) = \sum_{a,x}p^*(a|x,\lambda)p(x|y,\mu)p(g|a,y,\mu)
\end{equation}
so that \eqref{sm1} reads
\begin{multline}\label{sm2}
[1+I_R(\{\mathbb{M}_{x}\})]\sum_{\substack{\lambda g\mu \\ by}}q(b,y)p(\mu)\tr\left[\rho_{b|y}G^*_{\lambda}\right]p(g|\lambda,y,\mu)\delta_{b,g} \\
\geq \sum_{\mu abxyg}q(b,y)p(\mu)\tr\left[\rho_{b|y}M_{a|x}\right] \\ \times p(x|y,\mu)p(g|a,y,\mu)\delta_{b,g},
\end{multline}
The sum on the left hand side has the form of the success probability in QSD game with a single measurement, given in \eqref{e:psuc C}. It does not have the most general form, since $G^*_{\lambda}$ does not depend on $\mu$ (in this expression, $\lambda$ is playing the role of $a$ in \eqref{e:psuc C}). Hence, the sum is not larger than the optimal sucess probability with single measurement in the QSD game:
\begin{multline}\label{sm3}
[1+I_R(\{\mathbb{M}_x\})]P_g^{\JM}(\{\mathcal{E}_y\}) \\
\geq \sum_{\substack{\mu,a,b\\x,y,g}}q(b,y)p(\mu)p(x|y,\mu)\tr\left[\rho_{b|y}M_{a|x}\right]p(g|a,y,\mu)\delta_{b,g}.
\end{multline}
This expression holds for all $p(\mu)$, $p(x|y,\mu)$ and $p(g|a,y,\mu)$, so it must hold if we maximise both sides over all such probabilities (noting that the left hand side is in fact already independent of all of them):
\begin{multline}\label{sm5}
[1+I_R(\{\mathbb{M}_{x}\})]P_g^{\JM}(\{\mathcal{E}_y\}) \\
\geq \max_{\substack{p(\mu)\\ p(x|y,\mu)\\ p(g|a,y,\mu)}}\sum_{\mu abxyg}q(b,y)p(\mu)p(x|y,\mu) \times \\ \times\tr\left[\rho_{b|y}M_{a|x}\right]p(g|a,y,\mu)\delta_{b,g}.
\end{multline}
The right-hand-side is now equal to the optimal success in the QSD game with incompatible measurements as defined in \eqref{e:Pg}. This holds for all QSD games, (collections of ensembles $\{\mathcal{E}_y\}_y$. Thus, re-arranging and maximising over all games we arrive at the following inequality
\begin{equation}\label{e:ub}
1+I_R(\{\mathbb{M
}_{x}\}) \geq \max_{\{\mathcal{E}_y\}} \frac{P_g(\{\mathcal{E}_y\},\{\mathbb{M}_x\})}{P_g^{\JM}(\{\mathcal{E}_y\})}.
\end{equation}
This proves that upper bound, that $1 + I_R(\{\mathbb{M}_x\})$ is always larger than the advantage in any QSD game. 

\section{Lower bound}

In this section we now show that the upper bound from the previous section can be achieved, by exhibiting a carefully chosen optimal game $\{\mathcal{E}^*_y\}_y$, that has advantage equal to $1 + I_R(\{\mathbb{M}_x\})$ when played with $\{\mathbb{M}_x\}_x$.

 Consider the optimal dual variables $\omega^*_{ax}$ and $X^*$ from the dual SDP formulation of the RoI as defined in \eqref{dual}. Those variables satisfy
\begin{align} \nonumber
1 + I_R(\{\mathbb{M}_{x}\}) &= \tr\sum_{a,x}\omega^*_{ax}M_{a|x}, \\ \nonumber
\tr[X^*] &= 1,\qquad \omega^*_{ax} \geq 0\\
X^* &\geq \sum_{a,x}D_{\ab}(a|x)\omega^*_{a,x}, \qquad \forall \ab. \label{optDual}
\end{align}
Let us now introduce the following auxiliary variables
\begin{align*}
N^* &= \tr\sum_{a,x}\omega^*_{ax},\\
q^*(a,x) &= \frac{\tr\omega^*_{ax}}{N^*},\\
\rho^*_{a|x} = \frac{\omega^*_{ax}}{\tr\omega^*_{ax}} &= \frac{\omega^*_{ax}}{N^*q^*(a,x)}.
\end{align*}
The variables $\rho^*_{a|x}$ are normalised quantum states for all $a$, $x$ by construction, while $\{q^*(a,x)\}$ is a normalised probability distribution. By using the auxiliary variables the first constraint from \eqref{optDual} reduces to
\begin{equation}
1 + I_R(\{\mathbb{M}_{x}\}) = N^*\sum_{a,x}q^*(a,x)\tr\left[\rho^*_{a|x}M_{a|x}\right]
\end{equation}
Let us now assume that the QSD game is played with the set of ensembles $\{\mathcal{E}_y^*\}_y$, where $\mathcal{E}^*_y = \{q^*(b|y),\rho^*_{b|y}\}$,  $q^*(b|y) = q^*(b,y)/q^*(y)$, and $q^*(y) = \sum_b q^*(b,y)$ is the probability that Bob sends $y$ to Alice. The strategy for playing the game is taken to be the following:
\begin{itemize}
\item $p(\mu) = \delta_{\mu,0}$,
\item $p(x|y,\mu = 0) = \delta_{y,x}$, \textit{i.e.} we measure $\mathbb{M}_y$ when given $y$,
\item $p(g|a,y,\mu = 0) = \delta_{g,b}$, \textit{i.e.} we guess that $b=g=a$ when get outcome $a$.
\end{itemize}
The score achieved by this strategy is a lower bound on $P_g(\{\mathcal{E}_y^*\},\{\mathbb{M}_{x}\})$, (since this is a potentially sub-optimal strategy for playing). It therefore holds that
\begin{align}\nonumber
P_g(\{\mathcal{E}_y^*\},\{\mathbb{M}_{x}\}) &\geq \sum_{\substack{a,b,x\\y,g,\mu}}q^*(b,y)\delta_{\mu,0}\delta_{x,y}\tr\left[\rho^*_{b,y}M_{a|x}\right]\delta_{g,b}\delta_{a,g} \\ \nonumber
&= \sum_{a,x}q^*(a,x)\tr\left[\rho^*_{a|x}M_{a|x}\right]\\ \label{smineq1}
&= \frac{1}{N^*}(1 + I_R(\{\mathbb{M}_{x}\}))
\end{align}
As a short digression, which will be useful later, let us look more carefully at the strategies $P_g^{\JM}(\{\mathcal{E}_y\})$:
\begin{multline}
P_g^{\JM}(\{\mathcal{E}_y\}) = \max_{\substack{\mathbb{G}_{\nu}\\ p(g|y,\nu)\\p(\nu)}}\sum_{\substack{a,b,y\\ g,\nu}}q(b,y)p(\nu)\\ \times \tr\left[\rho_{b,y}G_{a|\nu}\right]p(g|a,y,\nu)\delta_{b,g}
\end{multline}
In in the first section of the appendix, one can decompose $p(g|a,y,\nu)$ into deterministic distributions. For that purpose introduce $D_{\mathbf{b}}(g|y) = \delta_{g,\mathrm{b}_x}$ to be functions such that $g$ is deterministically equal to $b_x$ where $\mathbf{b}$ is a string of outcomes, one for each measurement setting. It is always possible to write
\begin{equation}
p(g|a,y,\nu) = \sum_{\mathbf{b}}p(\mathbf{b}|a,\nu)D_{\mathbf{b}}(g|y)
\end{equation}
This decomposition allows one to obtain
\begin{equation}\label{sm6}
\begin{split}
\sum_{\substack{a,b,y\\g,\nu}}&q(b,y)p(\nu)\tr\left[\rho_{b|y}G_{a|\nu}\right]p(g|a,y,\nu)\delta_{b,g} \\ &= \sum_{b,y,\mathbf{b}}q(b,y)\tr\left[\rho_{b|y}\left(\sum_{a,\nu}p(\nu)G_{a|\nu}p(\mathbf{b}|a,\nu)\right)\right]\\ &\qquad \qquad \times D_{\mathbf{b}}(g|y)\delta_{b,g} \\ &=  \sum_{b,y,\mathbf{b}}q(b,y)\tr\left[\rho_{b,y}\tilde{G}_{\mathbf{b}}\right]D_{\mathbf{b}}(g|y)\delta_{b,g}
\end{split}
\end{equation}
where to obtain the third line we introduced the new variable
\begin{equation}
\tilde{G}_{\mathbf{b}} = \sum_{a,\nu}p(\nu)G_{a|\nu}p(\mathbf{b}|a,\nu)
\end{equation}
For all values of $\mathbf{b}$ this variable is positive semi-definite and it satisfies the following completeness relation
\begin{align}
\sum_{\mathbf{b}}\tilde{G}_{\mathbf{b}} &= \sum_{a,\mathbf{b},\nu}p(\nu)G_{a|\nu}p(\mathbf{b}|a,\nu)\nonumber \\
&= \sum_{a,\nu}p(\nu)G_{a|\nu}\nonumber\\
&= \sum_{\nu}p(\nu)\mathds{1}\nonumber\\
&= \mathds{1}
\end{align}
The second equality is a simple consequence of the fact that $p(\mathbf{b}|a,\nu)$ is a probability distribution, while the third one comes from the fact that $G_{a|\nu}$ is a valid measurement. Hence, positivity and completeness of $\tilde{G}_{\mathbf{b}}$ ensure that it represents a valid POVM. Eq. \eqref{sm6} means that we can, without loss of generality, assume that we measure $\rho_{b|y}$ in order to make a guess for every possible value of $b$ for each $y$ and later simply announce the value $g = b_y$ once we know $y$. The above shows that this is in fact as good as the most general strategy and thus
\begin{equation}
P_g^{\JM}(\{\mathcal{E}_y\}):= \max_{\tilde{\mathbb{G}}}\sum_{b,y,\mathbf{b}}q(b,y)\tr\left[\rho_{b,y}\tilde{G}_{\mathbf{b}}\right]D_{\mathbf{b}}(g|y)\delta_{b,g}
\end{equation}

Let us now return to the variable $N^*$. From the definition of $N^*$  and the dual SDP formulation it follows
\begin{equation*}
X^* \geq \sum_{b,y}D_{\mathbf{b}}(b|y)N^*q^*(b,y)\rho^*_{b|y}.
\end{equation*}
Multiplying by and arbitrary $\tilde{G}_{\mathbf{b}}$, summing over $\mathbf{b}$ and tracing leads to
\begin{equation}
\tr\sum_{\mathbf{b}}X^*\tilde{G}_{\mathbf{b}} \geq \sum_{b,y,\mathbf{b}}D_{\mathbf{b}}(g|y)\delta_{b,g}N^*q^*(b,y)\tr\left[\tilde{G}_{\mathbf{b}}\rho^*_{b,y}\right]
\end{equation}
Since $\tilde{G}_{\mathbf{b}}$ is a valid POVM  and $X^*$ has unit trace the left-hand-side of the inequality is equal to one. As it holds for all $\tilde{G}_{\mathbf{b}}$, it holds if the expression is maximized over $\tilde{G}_{\mathbf{b}}$, which implies
\begin{equation*}
\max_{\tilde{G}_{\mathbf{b}}}\frac{1}{N^*} \geq \max_{\tilde{G}_{\mathbf{b}}}\sum_{b,y,\mathbf{b}}q^*(b,y)\tr\left[\tilde{G}_{\mathbf{b}}\rho^*_{b,y}\right]D_{\mathbf{b}}(g|y)\delta_{b,g}.
\end{equation*}
This furthermore implies
\begin{equation}
\frac{1}{N^*} \geq P_g^{\JM}(\{\mathcal{E}_y^*\}).
\end{equation}
This inequality, together with \eqref{smineq1} implies
\begin{equation}
\frac{P_g(\{\mathcal{E}_y^*\},\{\mathbb{M}_x\})}{P_g^{\JM}(\{\mathcal{E}_y^*\}} \geq 1+I_R(\{\mathbb{M}_{x}\}).
\end{equation}

However, since we already proved in \eqref{e:ub} that $1+I_R(\{\mathbb{M}_{x}\})$ upper bounds the success probability for any QSD game $\{\mathcal{E}_y\}_y$, it must be the case that $\{\mathcal{E}^*_y\}_y$ is equal to $1+I_R(\{\mathbb{M}_{x}\})$, which completes the proof of the main result. 

\section{Monotones for measurement simulation}

In this section we prove that the measurements $\{\mathbb{M}_{x}\}_x$ can simulate another set of measurements $\{\mathbb{M}'_{y}\}_y$ if and only if $\{\mathbb{M}'_{y}\}_y$ never outperforms $\{\mathbb{M}_{x}\}_x$ in the QSD game introduced in the main text for every ensemble of states:
%
\begin{multline}\label{statement}
\{\mathbb{M}_{x}\} \succ \{\mathbb{M}'_{y}\} \\ \iff P_g(\{\mathcal{E}_z\},\{\mathbb{M}_x\}) \geq P_g(\{\mathcal{E}_z\},\{\mathbb{M}'_y\}) \\ \forall \{\mathcal{E}_z\}.
\end{multline} 
Recall that the success in the QSD game is defined as (we change notation here slightly, using $z$ and $c$ for the QSD game, as we will use $b$ and $y$ for the measurements $\{\mathbb{M}'_y\}$):
\begin{multline}
P_g(\{\mathcal{E}_z\},\{\mathbb{M}_x\}) = \max_{\substack{p(x|z,\mu)\\p(g|a,z,\mu)\\p(\mu)}}\sum_{\substack{a,c,g\\x,z,\mu}}q(c,z)p(\mu)p(x|z,\mu) \times \\ \times\tr\left[\rho_{c|z}M_{a|x}\right]p(g|a,z,\mu)\delta_{c,g}.
\end{multline}
By introducing a new set of measurements $\{\mathbb{M}'_{z}\}_z$, where $\mathbb{M}'_z = \{M_{g|z}\}_g$, which can be simulated by $\{\mathbb{M}
_{x}\}_x$ according to the definition of the simulation
\begin{equation}\label{smMon1}
M'_{g|z} = \sum_{a,x,\mu}p(\mu)p(x|y,\mu)M_{a|x}p(g|a,z,\mu)
\quad \forall b,y
\end{equation}
the success probability $P_g(\{\mathcal{E}_z\},\{\mathbb{M}_x\})$ can be re-expressed in a conceptually simpler form:
\begin{equation}\label{smMon2}
P_g(\{\mathcal{E}_z\},\{\mathbb{M}_x\}) = \max_{\{\mathbb{M}'_{z}\}\prec \{\mathbb{M}_{x}\}}\sum_{c,z,g}q(c,z)\tr\left[\rho_{c|z}M'_{g|z}\right]\delta_{c,g}
\end{equation}
That is, we see that the optimisation carried out can be thought of as optimising over all measurements $\{\mathbb{M}'_z\}_z$ that can be simulated by $\{\mathbb{M}_x\}_x$, where by definition now the outcome of the measurement is the guess $g$ of the corresponding state $c$ from the ensemble. 

Given this equivalent formulation, it is immediate that one direction of \eqref{statement} is immediately satisfied:
\begin{multline}
\{\mathbb{M}_{x}\} \succ \{\mathbb{M}'_{y}\}  \\ \implies  P_g(\{\mathcal{E}_z\},\{\mathbb{M}_x\}) \geq P_g(\{\mathcal{E}_z\},\{\mathbb{M}'_y\}) \quad \forall \{\mathcal{E}_z\}.
\end{multline}
Now we want to prove the converse direction. For that purpose assume $P_g(\{\mathcal{E}_z\},\{\mathbb{M}_x\}) - P_g(\{\mathcal{E}_z\},\{\mathbb{M}'_y\}) \geq 0$ for all QSD games $\{\mathcal{E}_z\}_z$. This assumption, written in full is
\begin{widetext}
\begin{multline}
\max_{\substack{p(x|z,\mu)\\p(g|a,z,\mu)\\p(\mu)}}\sum_{\substack{a,c,g\\x,z,\mu}}q(c,z)p(\mu)p(x|z,\mu)  \tr\left[\rho_{c|z}M_{a|x}\right]p(g|a,z,\mu)\delta_{b,g} \\ -\max_{\substack{p'(y|z,\nu)\\p'(g|b,z,\nu)\\p'(\nu)}}\sum_{\substack{b,c,g\\y,z,\nu}}q(c,z)p'(\nu)p'(y|z,\nu) \tr\left[\rho_{c|z}M'_{b|y}\right]p'(g|b,z,\nu)\delta_{c,g} \geq 0
\end{multline}
\end{widetext}
Let us now make a guess for a possibly sub-optimal strategy:
\begin{itemize}
\item $p'(nu) = \delta_{\nu,0}$,
\item $p'(y|z,\nu=0) = \delta_{y,z}$,
\item $p'(g|b,z,\nu=0) = \delta_{g,b}$.
\end{itemize}
This strategy implies
\begin{align}
\max_{\substack{p(x|z,\mu)\\p(c|a,z,\mu)\\p(\mu)}}&\sum_{\substack{a,c\\x,z,\mu}}q(c,z)p(\mu)p(x|z,\mu)  \tr\left[\rho_{c|z}M_{a|x}\right]p(c|a,z,\mu) \nonumber \\ -&\sum_{c,z}q(c,z)\tr\left[\rho_{c|z}M'_{c|z}\right] \geq 0
\end{align}
which after re-arranging gives 
\begin{multline}
\max_{\substack{p(x|z,\mu)\\p(c|a,z,\mu)\\p(\mu)}}\sum_{c,z}q(c,z)\tr\bigg[\rho_{c|z}\bigg(\sum_{a,x,\mu} p(\mu)p(x|z,\mu)\\ \times M_{a|x}p(c|a,z,\mu) - M'_{c|z}\bigg)\bigg] \geq 0
\end{multline}
This must be true for all $\{\mathcal{E}_z\}_z$, with $\mathcal{E}_z = \{q(c|z),\rho_{c|z}\}_c$. It therefore holds if minimised over all such QSD games:
\begin{multline}
\min_{\{\mathcal{E}_z\}}\max_{\substack{p(x|z,\mu)\\p(c|a,z,\mu)\\p(\mu)}}\sum_{c,z}q(c,z)\tr\bigg[\rho_{c|z}\bigg(\sum_{a,x,\mu} p(\mu)p(x|z,\mu)\\ \times M_{a|x}p(c|a,z,\mu) - M'_{c|z}\bigg)\bigg] \geq 0
\end{multline}
This expression is linear in $\{\mathcal{E}_z\}_z$, \textit{i.e.} in $\sigma_{c|z} = q(c,z)\rho_{c|z}$, which means also convex in these variables, and it is concave in $\{p(x|z,\mu),p(c|a,z,\mu),p(\mu)\}$. Therefore we can apply the minimax theorem \cite{neumann} and interchange the minimization and maximization. The last inequality, thus, reads
\begin{equation}
\max_{\substack{p(x|z,\mu)\\p(c|a,z,\mu)\\p(\mu)}}\min_{\{\mathcal{E}_z\}}\sum_{c,z}q(c,z)\tr\left[\rho_{c|z}\Delta_{cz}\right] \geq 0,
\end{equation}
where we have introduced
\begin{equation}
\Delta_{cz} = \sum_{a,x,\mu} p(\mu)p(x|z,\mu) M_{a|x}p(c|a,z,\mu) - M'_{c|z}
\end{equation}
Now, If $\{\mathbb{M}_{x}\} \succ \{\mathbb{M}'_{z}\}$, there exist $p(x|z,\mu),p(c|a,z,\mu)$ and $p(\mu)$ such that $\Delta_{cz} = 0$ for all values of $c$ and $z$. Let us assume that this is not true -- i.e. that no such $p(x|z,\mu),p(c|a,z,\mu)$ and $p(\mu)$ exist, in other words that $\Delta_{cz} \neq 0$ for all $c$ and $z$. In what follows we will show, by contradiction, that this is impossible. 

First, note that
\begin{align}
\sum_c\Delta_{cz} &= \sum_{a,c,x,\mu}p(\mu)p(x|z,\mu)M_{a|x}p(c|a,z,\mu) - \sum_cM'_{c|z} \nonumber \\
&= \sum_{a,x,\mu}p(\mu)p(x|z,\mu)M_{a|x} - \mathds{1}\nonumber\\
&= \sum_{x,\mu}p(\mu)p(x|z,\mu)\mathds{1} - \mathds{1} \\ &= 0
\end{align}
The second line is a consequence of the normalisation of $p(c|a,z,\mu)$ and completeness of each $\mathbb{M}'_{z}$. The third line follows from the completeness of each $\mathbb{M}_{x}$ and the last from from the normalisation of $p(\mu)$ and $p(x|z,\mu)$. Since $\sum_c\Delta_{cz} = 0$ it is impossible that $\Delta_{cz}\geq 0$ for all $c,z$, since this would only happen if all $\Delta_{cz}$ vanished identically, but by assumption this isn't the case.

Hence, for each $z$, there must be at least one $c^*(z)$ such that $\Delta_{c^*(z)z}$ has a negative eigenvalue. Let us denote by $\ket{\psi_{c^*(z)z}}$ the corresponding eigenvector with eigenvalue $\psi_{c^*(z)z} < 0$. Now let us choose $\{\mathcal{E}_z^*\}_z$ such that 
\begin{itemize}
\item $q^*(c,z) = q^*(c|z)q^*(z)$,
\item  $q^*(z) = 1/n$,
\item $q^*(c|z) = \delta_{c^*(z),c}$,
\item $\rho^*_{c^*(z)|z} = \ket{\psi_{c^*(z)z}}\bra{\psi_{c^*(z)|z}}$
\end{itemize}
Then 
\begin{equation}
\sum_{c,z}q^*(c,z)\tr\left[\rho^*_{c|z}\Delta_{cz}\right] = \frac{1}{n}\sum_z\psi_{c^*(z)z} < 0, 
\end{equation}
which is a contradiction, since by assumption  $\sum_{cz}q(c,z)\tr\left[\rho_{c|z}\Delta_{cz}\right] \geq 0$.
Therefore, there must exist $p(x|z,\mu),p(c|a,z,\mu)$ and $p(\mu)$ such that $\sum_{a,x,\mu} p(\mu)p(x|z,\mu) M_{a|x}p(c|a,z,\mu) = M'_{c|z}$ and hence $\{\mathbb{M}_{x}\} \succ \{\mathbb{M}'_{z}\}$. By this we have proven that 
\begin{multline}
P_g(\{\mathcal{E}_z\},\{\mathbb{M}_x\}) \geq P_g(\{\mathcal{E}_z\},\{\mathbb{M}'_y\}) \quad \forall \{\mathcal{E}_z\}   \\
\implies \{\mathbb{M}_{x}\} \succ \{\mathbb{M}'_{y}\}
\end{multline}
which together with the already proven converse statements implies \eqref{statement}.
In words, this shows that the guessing probabilities for all QSD games $\{\mathcal{E}_z\}_z$ constitute a complete set of monotones for the partial order $\{\mathbb{M}_{x}\} \succ \{\mathbb{M}'_{y}\}$.

Finally, let us show how this relates to the RoI. Assume $\{\mathbb{M}_x\}_x$ has optimal QSD game $\{\mathcal{E}^*_z\}_z$ such that $1+I_R(\{\mathbb{M}_x\}) = P_g(\{\mathcal{E}_z^*\},\{\mathbb{M}_x\})/P_g^{\JM}(\{\mathcal{E}^*_z\})$. Analogously, assume $\{\mathbb{M}'_y\}$ has the optimal game $\{\mathcal{F}^*_z\}_z$ such that $1+I_R(\{\mathbb{M}'_y\}) = P_g(\{\mathcal{F}^*_z\},\{\mathbb{M}'_y\})/P_g^{\JM}(\{\mathcal{F}^*_z\})$. Let us assume $\{\mathbb{M}_x\}\succ\{\mathbb{M}'_y\}$. Then
\begin{align*}
    1 + I_R(\{\mathbb{M}_x\}) &= \frac{P_g(\{\mathcal{E}^*_z\},\{\mathbb{M}_x\})}{P_g^{\JM}(\{\mathcal{E}^*_z\})} \\
    &\geq \frac{P_g(\{\mathcal{F}^*_z\},\{\mathbb{M}_x\})}{P_g^{\JM}(\{\mathcal{F}^*_z\})} \\
    &\geq \frac{P_g(\{\mathcal{F}^*_z\},\{\mathbb{M}'_y\})}{P_g^{\JM}(\{\mathcal{F}^*_z\})} \\
    &=  1 + I_R(\{\mathbb{M}'_y\})
\end{align*}
The first inequality follows from the fact that $\{\mathcal{E}^*_z\}_z$ is the optimal QSD game for $\{\mathbb{M}_x\}_x$. The second inequality follows from from \eqref{statement}. Thus we conclude that $I_R(\{\mathbb{M}_x\}) > I_R(\{\mathbb{M}'_y\})$ whenever $\{\mathbb{M}_{x}\} \succ \{\mathbb{M}'_{y}\}$, \textit{i.e.} the RoI is also a monotone for measurement simulation.

\end{appendix}

\end{document}